\documentclass[12pt]{iopart}
\usepackage{graphicx}
\usepackage{bm}
\usepackage{color}
\usepackage{ulem}
\usepackage{url}
\usepackage[colorlinks,linkcolor=blue,anchorcolor=blue,citecolor=blue,urlcolor=blue]{hyperref}
\usepackage{amssymb}

\begin{document}
\title[]{Highly efficient laser-driven Compton gamma-ray source}

\author{T. W. Huang$^{1,2,3}$, C. M. Kim$^{1,4}$, C. T. Zhou$^{2,3}$, M. H. Cho$^{1}$, K. Nakajima$^{1}$, C. M. Ryu$^{1}$, S. C. Ruan$^{2,3}$, and C. H. Nam$^{1,5}$}
\address{$^1$ Center for Relativistic Laser Science, Institute for Basic Science, Gwangju 61005, Republic of Korea}
\address{$^2$ Center for Advanced Material Diagnostic Technology, Shenzhen Technology University, Shenzhen 518118, People's Republic of China}
\address{$^3$ College of Optoelectronic Engineering, Shenzhen University, Shenzhen 518060, People's Republic of China}
\address{$^4$ Advanced Photonics Research Institute, Gwangju Institute of Science and Technology, Gwangju 61005, Republic of Korea}
\address{$^5$ Department of Physics and Photon Science, Gwangju Institute of Science and Technology, Gwangju 61005, Republic of Korea}

\ead{chulmin@gist.ac.kr}

\begin{abstract}
The recent advancement of high-intensity lasers has made all-optical Compton scattering become a promising way to produce ultra-short brilliant
$\gamma$-rays in an ultra-compact system. However, so far achieved Compton $\gamma$-ray sources are severely limited by low conversion efficiency
(lower than $10^{-5}$) and spectral intensity ($\sim10^{4}$ ${\rm photons/0.1\%BW}$). Here we present a highly efficient gamma photon emitter
obtained by irradiating a high-intensity laser pulse on a miniature plasma device consisting of a plasma lens and a plasma mirror. This concept
exploits strong spatiotemporal laser-shaping process and high-charge electron acceleration process in the plasma lens, as well as an efficient
nonlinear Compton scattering process enabled by the plasma mirror. Our full three-dimensional particle-in-cell simulations demonstrate that in this novel scheme,
brilliant $\gamma$-rays with very high conversion efficiency (higher than $10^{-2}$) and spectral intensity ($\sim10^{9}$ ${\rm photons/0.1\%BW}$)
can be achieved by employing currently available petawatt-class lasers with intensity of $10^{21}$ ${\rm W/cm^2}$. Such efficient and intense
$\gamma$-ray sources would find applications in wide-ranging areas.

\end{abstract}
\vspace{2pc}

\maketitle

\section{Introduction}
\label{Introduction}
The pursuit of compact $\gamma$-ray sources is motivated by many applications in fundamental science, industry, and medicine \cite{Corde,Albert}.
High-energy $\gamma$-ray radiation has become an immensely useful tool for probing hot dense matter \cite{Ben},
photonuclear spectroscopy \cite{Schreiber}, inspection of nuclear waste \cite{Jones}, material synthesis \cite{Seguchi}, and cancer therapy \cite{Weeks}.
With the rapid development of laser technology, the production of $\gamma$-rays based on laser-plasma interactions has attracted considerable
attention in the past decade. In contrast to conventional $\gamma$-ray sources based on large-scale and costly particle
accelerators \cite{Ballam,Weller}, the $\gamma$-ray source based on an all-optical approach is much more compact because in laser-plasma interaction
process, extremely large accelerating fields \--- above $100$ ${\rm GV/m}$ \--- can be produced and lead to high-energy electron beams in an ultrashort
distance. In addition, laser-based $\gamma$-ray source possesses unique properties, such as ultrashort duration, ultrahigh brilliance, and small source
size \cite{Corde,Albert}, which potentially makes it possible to realize a tabletop $\gamma$-ray source with much higher spatiotemporal resolution than
the conventional ones.

Several schemes have been proposed to produce compact $\gamma$-rays based on laser-plasma interactions, including
laser-driven bremsstrahlung radiation \cite{Kmetec,Gahn,Edwards,Glinec,Giulietti,Cipiccia} and Compton (or Thomson)
scattering \cite{Schwoerer,Powers,Chen,Sarri,Khrennikov,Yan,Phuoc,Tsai,Dopp,Yu}. For a laser-driven bremsstrahlung source,
where a high-energy electron beam, accelerated in the wakefield induced by an ultra-short and intense laser pulse as it propagates
in an underdense plasma, impinges on a target composed of high atomic number ($Z$) materials, very high-energy (up to 100 ${\rm MeV}$)
gamma photons can be generated \cite{Glinec,Giulietti,Cipiccia}. So far the peak brilliance of the $\gamma$-ray source
($\sim10^{17}$ ${\rm photons/s/mm^2/mrad^2/0.1\% BW}$) achieved in this scheme, however, is limited by wide divergence angle and
large source size, and the conversion efficiency (less than $10^{-3}$) is limited by a small cross section. The Compton (or Thomson)
scattering, based on the collision between a laser-wakefield accelerated electron beam and another counter-propagating laser
pulse \cite{Schwoerer,Powers,Chen,Sarri,Khrennikov,Yan} or a reflected laser pulse
by a plasma mirror \cite{Phuoc,Tsai,Dopp,Yu}, has been considered a promising scheme for the production of high-energy
high-brilliance $\gamma$-rays, since it exploits the double-Doppler upshift of the laser photon energy by relativistic electrons.
In this scheme, multi-${\rm MeV}$ $\gamma$-rays with high peak brilliance ($\sim10^{22}$ ${\rm photons/s/mm^2/mrad^2/0.1\% BW}$)
have already been produced experimentally \cite{Yu}. However, due to the limitations on the number
of high-energy electrons and the laser photons participating the scattering process, the presently achieved energy conversion efficiency (less than $10^{-5}$
of total laser energy) and the spectral intensity (less than $10^{5}$ ${\rm photons/0.1\%BW}$) of the $\gamma$-rays are rather
low \cite{Schwoerer,Powers,Chen,Sarri,Khrennikov,Yan,Phuoc,Tsai,Dopp,Yu} in this scheme, which severely limits their usefulness in aforementioned
applications.

Recently synchrotron photon emission in a relativistic transparency regime \cite{Zhidkov,Brady,Ji,Stark,Huang0,Chang},
where an ultra-intense laser pulse is incident onto a relativistically underdense target, was proposed as a relatively efficient scheme for
the production of brilliant $\gamma$-rays. The strong coupling of an intense laser pulse with a relativistically underdense plasma
can easily generate high-current high-energy electron beams and thus radiate high-flux $\gamma$-rays \cite{Huang}. However, this
scheme becomes effective for lasers at intensity exceeding $10^{22}$ ${\rm W/cm^2}$ or even $10^{23}$ ${\rm W/cm^2}$, which has good prospects
for the next-generation lasers \cite{Powell} but holds little promise for the presently available lasers \cite{Danson}. So far the production of
brilliant $\gamma$-rays, especially with high conversion efficiency and spectral intensity, which are the most essential parameters for
practical applications, however, remains out of reach.

Based on the very recent technological advances in high-power short-pulse laser with a very high contrast \cite{Kim,Wagner,Sung}, in this
article, we propose a highly efficient approach to produce copious gamma photons by irradiating a high-intensity laser pulse on a microsized
bilayer plasma device, which consists of a near-critical-density (NCD) plasma lens \cite{Ren,Wang,Bin} and a solid-density plasma mirror \cite{Thaury}.
The plasma lens acts to strongly compress the laser pulse and simultaneously provides the source for
high-charge (tens of ${\rm nC}$) and high-energy (hundreds of ${\rm MeV}$) electrons, which are directly accelerated by the enhanced
laser field ($\sim100$ ${\rm TV/m}$) within tens of micrometers. The plasma mirror acts to effectively reflect the laser pulse, and the reflected laser pulse
collides with the counter-propagating electron beam. In this case, nonlinear Compton scattering \cite{Sarachik,Bula} of the strongly focused
laser pulse by the high-charge high-energy electrons occurs, emitting copious gamma photons. As will be demonstrated,
this novel approach makes it feasible to produce brilliant $\gamma$-rays with very high conversion efficiency and spectral
intensity using currently available lasers at intensities $\sim10^{21}$ ${\rm W/cm^2}$.

The paper is organized as follows: section \ref{sec2} introduces the schematic of the efficient gamma photon emitter and
section \ref{sec3} introduces the simulation setup. In section \ref{sec4}, we describe the strong laser-shaping process and
high-charge electron acceleration process in NCD plasma lens. In section \ref{sec5}, the nonlinear Compton scattering process
induced by the plasma mirror is discussed and the properties of the emitted gamma photons are described. In section \ref{sec6},
we discuss the dependence of gamma-ray emission on laser intensity, plasma density, and also the length of the plasma lens. The
comparisons with single-layer target cases with the plasma lens alone or with the plasma mirror alone are also given.
In section \ref{sec7}, the significance of the results and comparison with previous works are discussed. The potential applications
of the intense gamma-ray source generated in present scheme are presented. The effects of Bremsstrahlung radiation and the prospects
of the present scheme for future lasers are also discussed. The summary is given in section \ref{sec8}.

\begin{figure}
\begin{center}
\includegraphics[width=12cm]{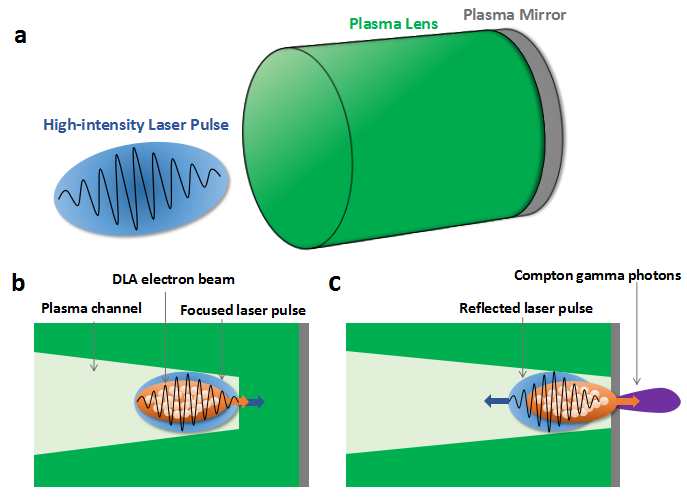}
\end{center}
\caption{{\bf Schematic of the gamma photon emitter}.
(a) A high-intensity laser pulse is incident onto a bilayer plasma device consisting of a plasma lens and a plasma mirror.
(b) At an early stage, the laser pulse (blue color) is strongly focused by the plasma lens (green color), and, meanwhile, a significant
number of electrons (orange color) are confined in the laser-produced plasma channel (light-green color), and they are directly accelerated
by the strong laser field. (c) At a later stage, the focused laser pulse (blue color) is reflected by the plasma mirror (grey color) and collides with
the forward-propagating electron beam. Then the nonlinear Compton scattering process is induced and copious gamma photons (purple color)
are emitted.
}\label{fig1}
\end{figure}

\section{Schematic of the gamma photon emitter}
\label{sec2}
The schematic of the efficient gamma photon emitter is shown in Fig.\ \ref{fig1}. In this approach, a laser pulse is incident onto
a microsized bilayer plasma device, as shown in Fig.\ \ref{fig1}(a), which consists of a plasma lens target with near-critical density
($0.1n_c<n_e<\gamma_0 n_c$) \cite{Robinson} and a plasma mirror target with solid density ($n_e\gg \gamma_0n_c$),
where $n_e$ is the plasma density, $n_c=\epsilon_0m_e\omega_L^2/e^2$ is the critical plasma density for laser propagation
in non-relativistic case, $\gamma_0=\sqrt{1+\langle a_0^2\rangle}$ is the average Lorentz factor, $a_0=eE/m_e\omega_Lc$ is
the dimensionless amplitude of laser electric field, $\epsilon_0$ is the vacuum permittivity, $m_e$ is the mass of electron,
$e$ is the elementary charge, $c$ is the speed of light in vacuum, and $\omega_L$ is the central laser frequency.
In this setup, the NCD plasma lens becomes transparent for the ultra-intense laser pulse ($a_0\gg1$) because of the relativistic
effect of electron motion in laser field \cite{Kaw}, which induces an increase in electron mass by the Lorentz factor
and thus an effectively reduced plasma density $n_e/\gamma_0$ in the relativistically corrected plasma refractive index, $n=\sqrt{1-n_e/\gamma_0 n_c}$.
The overdense plasma mirror is of sufficiently high plasma density and is still opaque for the laser pulse, thus it will reflect
the laser pulse when incident on the target.

This simple target design makes it feasible to achieve an efficient nonlinear Compton scattering process through the
combination of relativistic plasma optics and the compact laser plasma-based accelerators. The basic laser-plasma interaction
process can be seen in Figs.\ \ref{fig1}(b) and (c). At an early stage, the laser pulse
will penetrate into the plasma lens, as shown in Fig.\ \ref{fig1}(b). The plasma lens here has dual functions. On one hand,
the laser pulse undergoes rapid self-shaping process in the plasma lens due to the combined nonlinear effects of
relativistic self-focusing and relativistic self-phase modulation \cite{Ren,Wang,Bin}, which result from the modification
of plasma refractive index by the relativistic effect. 
As a result, a much shorter and stronger laser pulse is generated and the peak laser intensity can be significantly enhanced
during the shaping process. On the other hand, the ponderomotive force of the shaped laser pulse expels electrons from the
central region to form a plasma density channel, as shown in Fig.\ \ref{fig1}(b). In the plasma channel, still a significant number of electrons
are confined by the self-generated electromagnetic fields, and they are directly accelerated by the enhanced laser electric field in
the direct-laser acceleration (DLA) regime \cite{Pukhov}. Due to the relatively high plasma density and high-gradient accelerating
electric field, a high-charge high-energy electron beam can be produced in the plasma lens, and it moves together with the laser
pulse, as shown in Fig.\ \ref{fig1}(b). At a later stage, when the laser pulse reaches the solid-density target, its rising edge ionizes the target,
forming a plasma mirror \cite{Thaury} that effectively reflects the main part of the laser pulse, as shown in Fig.\ \ref{fig1}(c). In such case, the
reflected laser pulse collides with the forward-propagating electron beam. The nonlinear Compton scattering process between the high-charge
high-energy electron beam and the strongly focused laser pulse occurs, and copious gamma photons are emitted, as shown in Fig.\ \ref{fig1}(c).

The current setup provides a simple and efficient approach for gamma photon emissions. In this regime, both the number of high-energy electrons and the laser
photons participating the scattering process can be significantly increased. In particular, the enhanced laser photon
density (or laser intensity) leads to a dominant high-order multi-photon scattering process \cite{Sarachik,Bula}, which means that multiple photons can be
scattered from a single electron, and a single high-energy gamma photon is emitted. Note that, this scheme is different from
the well-established Compton scattering schemes based on laser-wakefield acceleration in rather low-density plasmas ($n_e\ll0.1n_c$), which are severely
limited by the number of high-energy electrons and laser photons, and provide inefficient gamma-ray sources.

\section{Simulation setup}
\label{sec3}
To demonstrate the efficiency of this novel gamma photon emitter, fully three-dimensional particle-in-cell (PIC) simulations were
conducted using the \textsc{EPOCH} code \cite{Arber} that incorporates the quantum photon emission process via a Monte Carlo algorithm \cite{Ridgers,Duclous}.
The radiation reaction effect on the electron dynamics is calculated self-consistently. In our simulations, the laser pulse is linearly
polarized and the peak intensity is $I_0=5.3\times 10^{21}$ ${\rm W/cm^2}$, which corresponds to $a_0=50$ for the laser pulse with a central
wavelength of $\lambda_L=0.8$ $\rm{\ \mu m}$. Here $a_0$ refers to the dimensionless vector potential of the laser pulse. The laser pulse
is characterized by a spatially Gaussian profile $a=a_0\exp(-r^2/\sigma^2)$ with $\sigma=8.8\lambda_L$ and a temporally Gaussian profile
with duration of $\tau=28$ $\rm{\ fs}$ (full width at half maximum, FWHM). Such laser pulse can be achieved by the femtosecond petawatt lasers
at CoReLS \cite{Sung}. In the simulations, the plasma target is composed
of the near-critical density (NCD) plasma layer and a thin solid-density aluminum plasma layer. For the NCD layer, the electron density
is $n_c=1.74\times 10^{27}$ ${\rm m^{-3}}$ and the thickness is $40\lambda_L$. For the solid-density layer, aluminum is initialized as $\mathrm{Al}^{11+}$ with
the electron density of $660n_c$ and the thickness is $1.25\lambda_L$. The simulation box is
$62.5\ \lambda_L\times 30\ \lambda_L\times 30\ \lambda_L$ in $x \times y \times z$ space with a cell size of $10$ $\rm{nm}$ along $x$ direction
and $80$ $\rm{nm}$ along $y$ and $z$ directions, respectively. Here $x$-axis corresponds to the laser propagation direction and $y$-axis corresponds
to the laser polarization direction. The simulation box is filled in with about $1.8\times 10^9$ macroparticles. The initial electron temperature is
assumed to be $1$ ${\rm keV}$. As a comparison, the simulation for the single target case without the second layer is also conducted.
In these simulations, current smoothing technique and high-order weighting functions are employed to suppress the numerical heating.

\section{Laser-shaping and electron acceleration in plasma lens}
\label{sec4}
\begin{figure}
\begin{center}
\includegraphics[width=12cm]{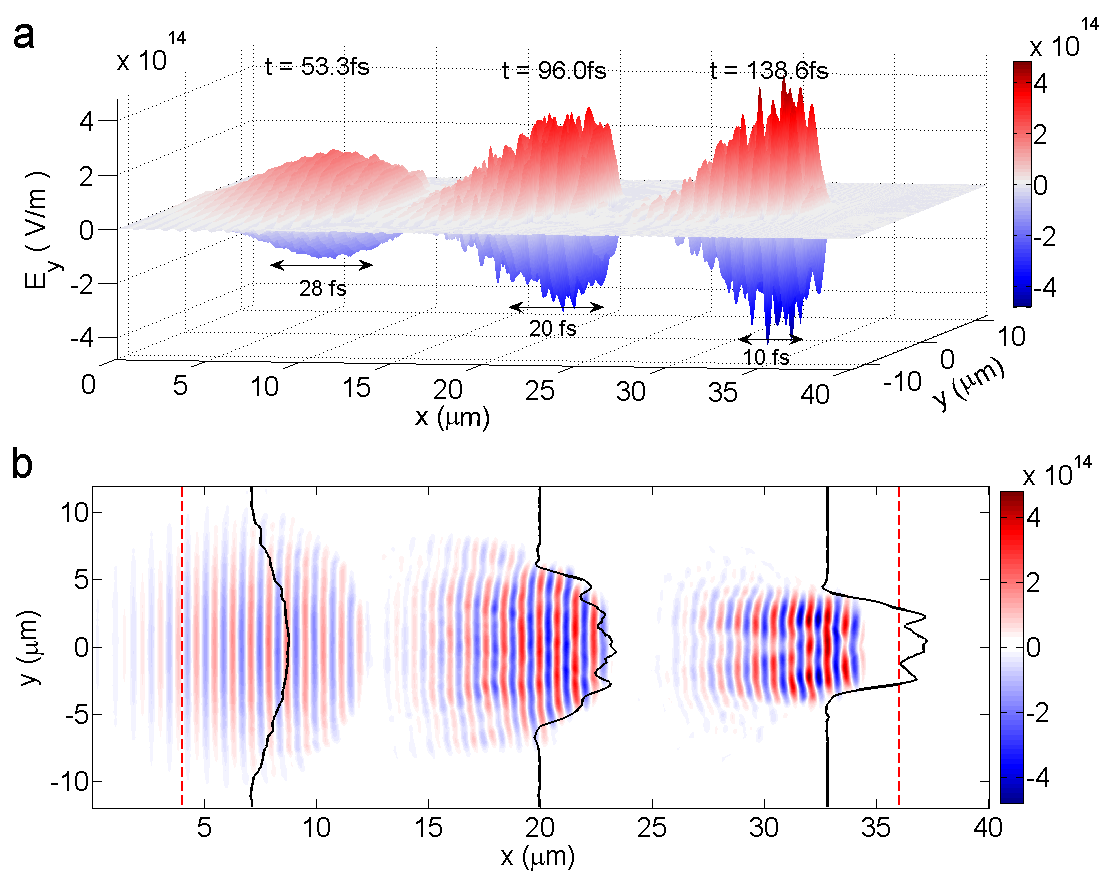}
\end{center}
\caption{{\bf Strong laser shaping process induced by the plasma lens.}
Time evolutions of the laser electric field ($E_y$, in units of ${\rm V/m}$) in plasma lens before the laser reflection.
(a) The spatial distribution of the transverse laser electric field at different times along the plane of $z=0$.
(b) The corresponding projection of (a). The vertical red-dashed lines in (b) mark the boundaries of the plasma lens and
the black-solid lines show the transverse profiles of the laser electric field at different positions.
}\label{fig2}
\end{figure}

The corresponding three-dimensional PIC simulation results are shown in Figs.2-5.
Due to the relativistic transparency, the laser pulse propagates into the plasma lens and experiences spatiotemporal self-shaping process,
as shown in Figs.\ \ref{fig2}(a) and (b). The corresponding self-focusing length can be estimated as \cite{Wang,Bin} $\sqrt{\ln{2}}\sigma(a_0n_c/n_e)^{1/2}$, where $\sigma$
is the beam radius. When the length of the plasma lens is comparable to this characteristic length (about $40$ ${\rm \mu m}$), the laser-shaping
process can occur effectively and the laser intensity can be remarkably enhanced. During the laser-shaping process, the laser beam gradually
shrinks into a radius about $2\sqrt{a_f}c/\omega_p$, which corresponds to the balance between the laser ponderomotive force and the radial
electrostatic force \cite{Huang2015}. Here $a_f$ refers to the dimensionless electric field of the self-shaped laser pulse and $\omega_p$ is the plasma frequency. Accordingly, the amplification factor of the laser intensity can be expressed as
$\frac{I_f}{I_0}\approx \left(\frac{n_e}{a_0n_c}\right)^{2/3}\left(\frac{\pi \sigma}{\lambda_L}\right)^{4/3}$,
where $I_f$ is the peak laser intensity of the self-shaped laser pulse, $I_0$ is the initial laser intensity, and $\lambda_L$ is the laser wavelength.
Figure \ref{fig2}(a) shows that, after the shaping process, the laser pulse duration (initially $28$ ${\rm fs}$) is reduced into about $10$ ${\rm fs}$
(FWHM). The peak laser intensity is enhanced by a factor of $6$ and reaches about $3.2\times10^{22}$ ${\rm W/cm^2}$,
as indicated from Fig.\ \ref{fig2}(b), which is in good agreement with the theoretical amplification factor of $6.1$. This clearly demonstrates that the
relativistic plasma lens can induce strong shaping for the laser pulse, leading to the generation of a much shorter and stronger laser pulse.
The laser focusing induced by the plasma lens is more effective than the optical focusing by the reflection of a plasma mirror \cite{Tsai2}, where
the latter is widely employed in previously proposed schemes. The comparison of these two focusing regimes shall be discussed later. In our case,
the necessary level of the input laser intensity for efficient gamma photon emissions could be relaxed significantly through the relativistic
self-shaping process.

\begin{figure}
\begin{center}
\includegraphics[width=12cm]{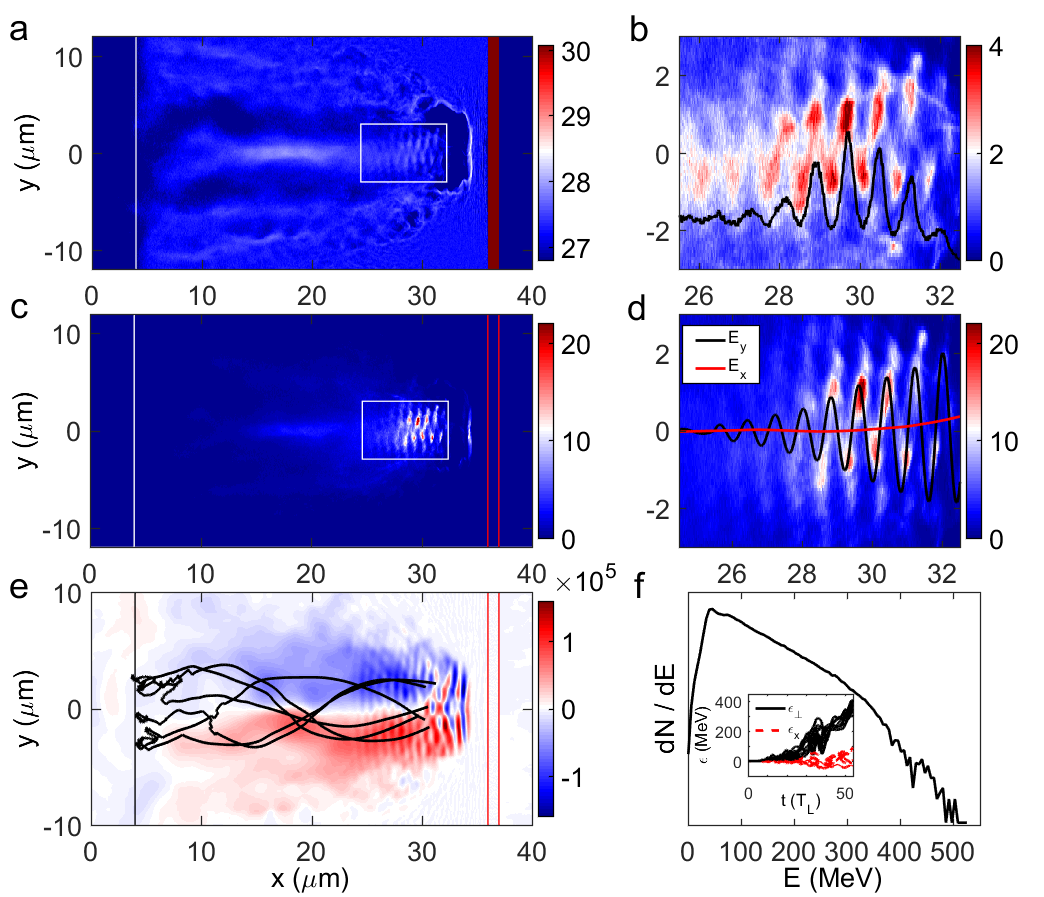}
\end{center}
\caption{{\bf High-charge electron acceleration in plasma lens.}
The transverse distribution of electrons and the self-generated magnetic field at $t=52T_L$ (before the laser reflection) along the plane
of $z=0$, where $T_L$ refers to the laser cycle.
(a) The electron density distribution in $\log_{10}$ scale ($\log_{10}n_e$, in units of ${\rm m^{-3}}$).
(b) The normalized electron density ($n_e/n_c$) distribution in the white rectangular area of (a).
The black curve in (b) shows the corresponding longitudinal profile at $y=1$ ${\rm \mu m}$.
(c) The normalized energy density ($E_en_e/100m_ec^2n_c$) distribution of the electrons, where $E_e$ refers to the average electron energy on each grid.
(d) Enlargement of the white rectangular area in (c). The black curve and red curve in (d) show the profiles
of the transverse laser electric field ($E_y$) and the longitudinal charge separation electric field ($E_x$) along $y=0$ ${\rm \mu m}$, respectively.
(e) Distribution of the azimuthal magnetic field $B_z$ (in units of ${\rm T}$). The black curves in (e) show the typical electron trajectories
in (x,y) plane. (f) The energy spectrum (number of electrons in $5$ ${\rm MeV}$ energy bandwidth) of the electrons located in a cylindrical volume
that corresponds to the white rectangular section in (a) and (c). The inset figure in (f) shows the time evolution of energy gain of the tracing electrons
in longitudinal direction ($\epsilon_x$) and in perpendicular direction ($\epsilon_\perp$), where $\epsilon_x=-\int{ev_xE_xdt}$ and $\epsilon_\perp=-\int{ev_yE_ydt}$
refer to the electron energy gain from the longitudinal charge separation electric field and the transverse laser field, respectively. The vertical
white lines in (a) and (c) and the vertical black lines in (e) mark the boundary of the plasma lens. The vertical red lines in (c) and (e) mark the boundaries
of the plasma mirror.
}\label{fig3}
\end{figure}

On the other hand, when the laser pulse propagates into the plasma lens, a plasma density channel is formed by the laser
ponderomotive force, as shown in Fig.\ \ref{fig3}(a). Figure \ref{fig3}(b) shows that a dense ($\sim5.6\times 10^{27}$ ${\rm m}^{-3}$) electron beam,
which is micro-bunched with the laser wavelength, is confined by the strong magnetic field ($\sim10^5$ ${\rm T}$)
self-generated in the plasma channel, as shown in Fig. \ref{fig3}(e). The particle tracing demonstrates that after an initial injection process,
electrons are confined by the magnetic field and execute betatron oscillations, as shown in Fig. \ref{fig3}(e). In the
present scheme, these electrons are directly accelerated by the strong laser field. The longitudinal charge separation electric field ($E_x$) is
negligible compared with the ultra-intense laser electric field ($E_y>10^{14}$ ${\rm V/m}$), as indicated in Fig. \ref{fig3}(d). To clarify the source
of energy gain of these electrons, the energy gain of each electron is separated into two parts, i.e., $\epsilon=\epsilon_x+\epsilon_\perp$,
where $\epsilon_x=-\int{ev_xE_xdt}$ and $\epsilon_\perp=-\int{ev_yE_ydt}$ refer to the electron energy gain from the longitudinal charge separation
electric field ($E_x$) and the transverse laser electric field ($E_y$), respectively. It is clearly shown from the inset of Fig. \ref{fig3}(f) that the
electrons are mainly accelerated by the laser electric field and the longitudinal charge separation electric field contributes little on the electron
acceleration. The laser electric field can accelerate the electrons to about $500$ ${\rm MeV}$ within tens of micrometers, as shown in Fig. \ref{fig3}(f).
As a result, an electron beam with ultra-high energy density ($\sim 10^{14}$ ${\rm J/m^3}$) is generated, as shown in Fig. \ref{fig3}(c).
These high-energy electrons co-propagate with the laser pulse and they are phase-locked to the laser electric field, as indicated in Fig. \ref{fig3}(d).
Figure \ref{fig3}(f) shows that the accelerated electron beam has a broad energy spectrum. The total charge of the electron beam reaches up
to $60$ ${\rm nC}$ (the charge of electrons with energy above $100$ ${\rm MeV}$ is about $26$ ${\rm nC}$), which is about three orders of magnitude
larger than that in previous Compton scattering scheme based on laser-wakefield electron accelerators using rather low-density plasmas \cite{Powers,Chen,Sarri,Khrennikov,Yan,Phuoc,Tsai,Dopp,Yu}. In addition, about $30\%$ of the laser energy is converted into such electron beam.
It clearly shows that the plasma lens, on the other hand, acts as an effective medium for high-charge (tens of ${\rm nC}$) and high-energy (hundreds of ${\rm MeV}$)
electron acceleration.

\section{Nonlinear Compton scattering and efficient gamma photon emissions}
\label{sec5}
\begin{figure}
\centering
\noindent\makebox[\textwidth][c]{\includegraphics[width=0.5\paperwidth]{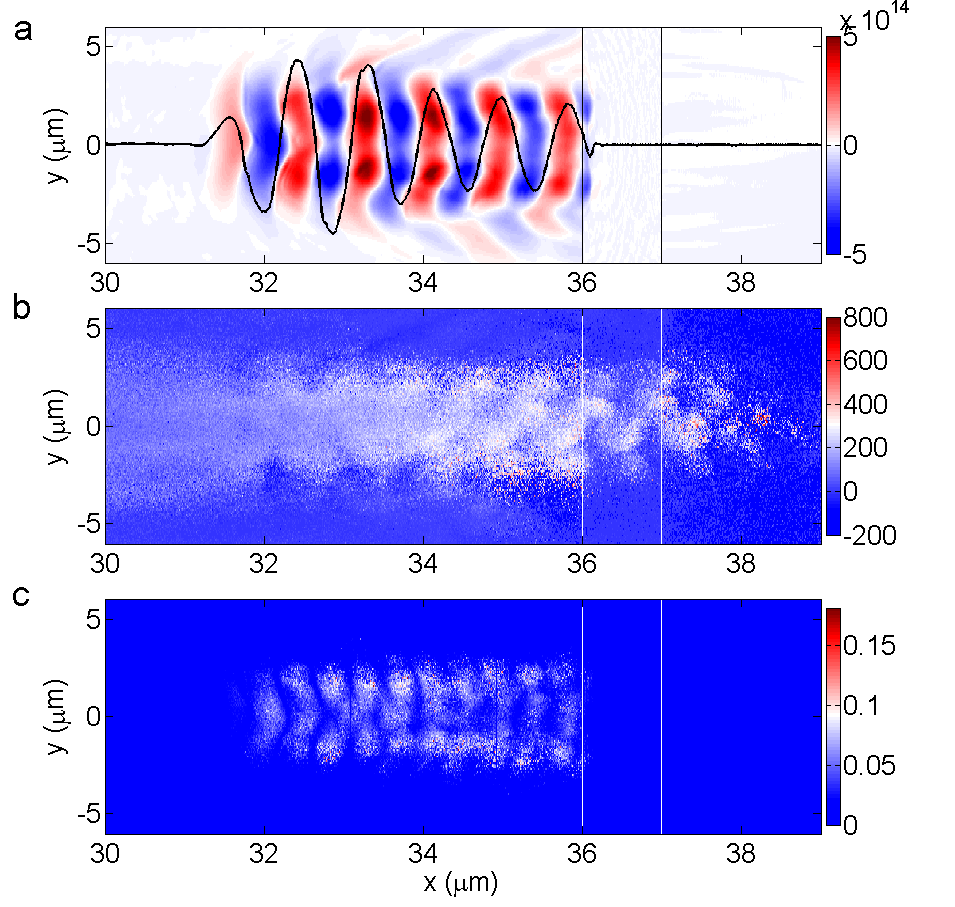}}
\caption{{\bf Laser-electron beam collision induced by the plasma mirror.}
The transverse distribution of the laser field and the electron beam at $t=60T_L$ (during the laser reflection) along the plane of $z=0$.
(a) Distribution of the reflected laser field ($(E_y-cB_z)/2$, in units of ${\rm V/m}$).
The black curve in (a) shows the corresponding longitudinal profile at $y=0$ ${\rm \mu m}$.
(b) Distribution of the longitudinal momentum ($p_x/m_ec$) of the electron beam.
(c) Distribution of the quantum photon emission parameter $\eta$ (averaged on each grid) for the electrons in (b).
The vertical black lines in (a) and the vertical white lines in (b) and (c) mark the boundary of the plasma mirror.
}\label{fig4}
\end{figure}

The laser-electron beam collision process enabled by the plasma mirror leads to efficient emission of gamma photons through nonlinear
Compton scattering, as shown in Fig.\ \ref{fig4}. The plasma mirror effectively reflects the enhanced laser pulse, and the reflected laser pulse
collides with the forward-moving electron beam, as indicated in Figs.\ \ref{fig4}(a) and (b). It is noted that the self-shaped laser pulse in plasma
lens is further focused after the reflection by the plasma mirror due to the relativistically curved surface \cite{Tsai2}, as can be seen from
Fig.\ \ref{fig2}(b) and Fig.\ \ref{fig4}(b), which induces an increase on the laser intensity from $3.2\times10^{22}$ ${\rm W/cm^2}$ to $3.8\times10^{22}$ ${\rm W/cm^2}$.
Despite this, in our scheme the relativistic self-focusing in plasma lens is still the dominant laser focusing scheme, which is more effective than the optical
focusing induced by plasma mirror. In the colliding process, the photons of the ultra-intense laser pulse ($a_f\approx 120$) are frequency-upshifted to $\gamma$-ray range
through nonlinear Compton scattering by the high-charge high-energy electrons. In this process, the quantum photon emission parameter \cite{Ridgers,Duclous} $\eta=\frac{\gamma_e}{E_s}\sqrt{\left(\mathbf{E}+\frac{1}{c}[\mathbf{v_e}\times\mathbf{B}]\right)^2-\frac{1}{c^2}(\mathbf{E}\cdot\mathbf{v_e})^2}$
is also calculated, as shown in Fig.\ \ref{fig4}(c), where $E_s=\frac{m_e^2c^3}{e\hbar}$ is the Schwinger field, $\hbar$ is the reduced Planck constant, $\mathbf{E}$
and $\mathbf{B}$ refer to the electric field and magnetic field acting on the electron, $\mathbf{v_e}$ is the electron velocity, and $\gamma_e$ is the Lorentz
factor. This parameter determines the photon emission rate and the radiation power by a single electron. In particular, when $\eta > 0.1$, the quantum effects
in the emission process become non-negligible \cite{Ridgers}. Figure \ref{fig4}(c) shows that the electron beam has a large average $\eta$ in the localized region
where the reflected laser pulse is present. Before the laser reflection, the maximum value of $\eta$ is only about $0.04$ in the copropagating configuration, where
the Lorentz force from the magnetic field of the laser pulse counteracts the force from the electric field of the laser pulse. After the laser reflection, the maximum
value of $\eta$ reaches about $0.4$, which is comparable with that achieved in previous experiment at SLAC \cite{Bula}. In the head-on colliding configuration, $\eta$ can
be simplified as $\eta\approx 2\gamma_e\frac{E_{\perp}}{E_s}\approx\hbar\omega_L\gamma_ea_f/m_ec^2\approx 2\varepsilon_{\gamma}/\varepsilon_e$, which compares the radiated
gamma photon energy to the electron energy \cite{Thomas}. Here $E_{\perp}$ refers to the electric field perpendicular to the electron velocity, $\hbar\omega_L$ is the laser
photon energy, $\varepsilon_{\gamma}\approx \hbar\omega_0a_f\gamma_e^2$ is the radiated photon energy, and $\varepsilon_e=\gamma_em_ec^2$ is the electron energy. In our case
with $\eta=0.4$, the quantum recoil effect is not negligible, thus the interaction process here is described as Compton scattering rather than Thomson scattering \cite{Piazza}. For electrons with $\varepsilon_e=500$ ${\rm MeV}$ and $\eta=0.4$, the energy of the radiated gamma photons can reach about $100$ ${\rm MeV}$.

\begin{figure}
\centering
\noindent\makebox[\textwidth][c]{\includegraphics[width=0.8\paperwidth]{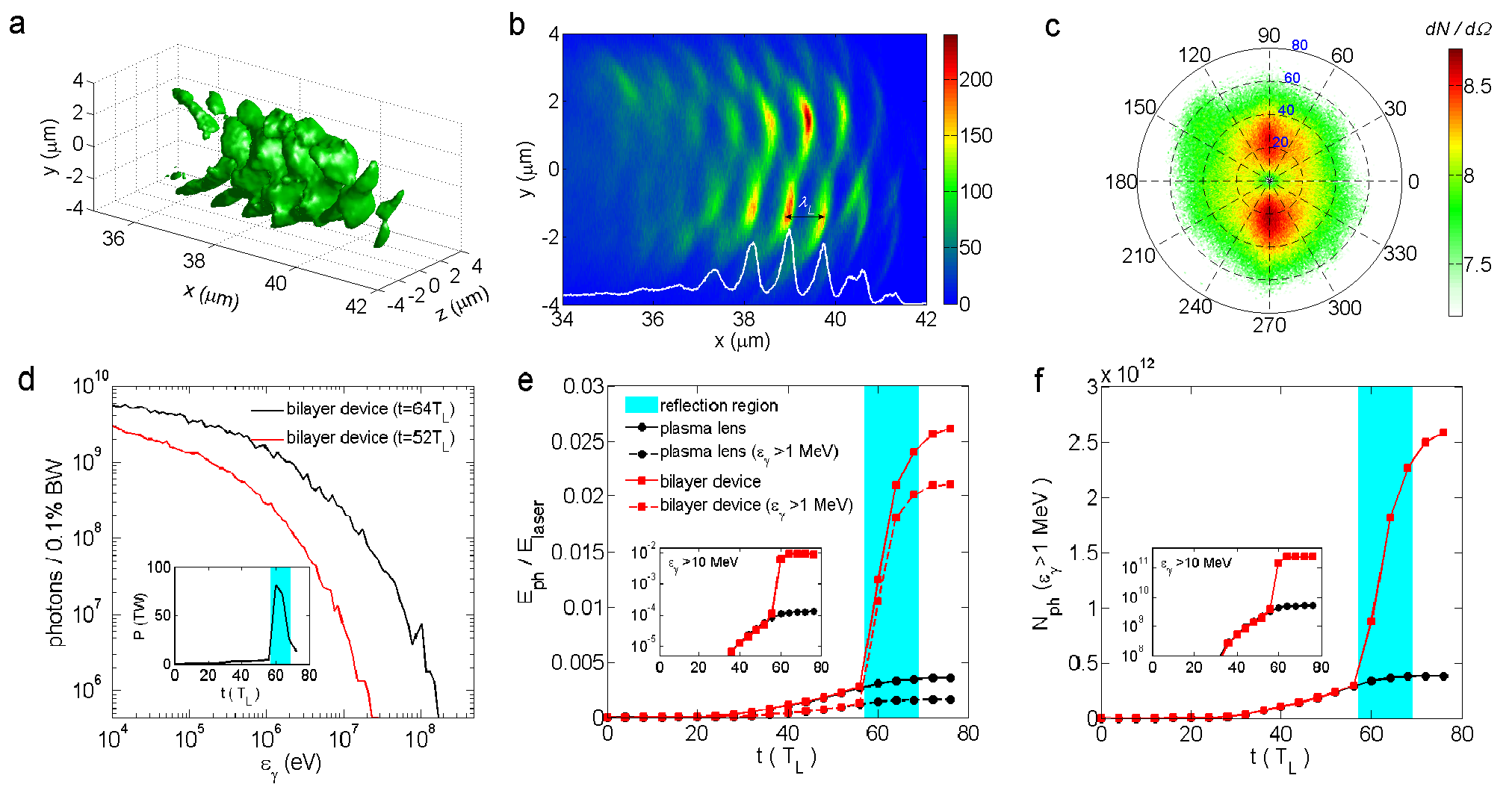}}
\caption{{\bf Copious gamma photon emissions.}
The properties of the emitted gamma photons at $t=64T_L$ (after the laser reflection), where only the forward-moving photons ($p_x>0$) are calculated.
(a) 3D isosurface distribution of the normalized energy density ($\varepsilon_\gamma n_\gamma/m_ec^2n_c$) of the gamma photons, where $\varepsilon_\gamma$
and $n_\gamma$ refer to the photon energy and photon density averaged on each grid, respectively.
(b) The corresponding transverse distribution of the energy density of the gamma photons. The white curve in (b) shows the longitudinal profile at $y=-1.2$ ${\rm \mu m}$.
(c) The angular distribution of the high-energy gamma photons ($\geq 1$ ${\rm MeV}$), where $d\Omega=\sin\theta d\theta d\phi$ is the solid angle,
$\theta=\arctan(p_{\perp}/p_x)$, $\phi=\arctan(p_y/p_z)$, and $p_{\perp}=\sqrt{p_y^2+p_z^2}$. The colorbar in (c) shows the number of gamma photons
emitted per square degree in $\log_{10}$ scale.
(d) The spectral intensity of gamma photons at $t=52T_L$ and $t=64T_L$. The inset figure in (d) shows the time evolution
of the total radiation power $P$ (in units of ${\rm TW}$), which is defined by the radiation energy emitted per laser cycle.
(e) Time evolutions of the conversion efficiency ($E_{ph}/E_{laser}$) of gamma photon emissions, where $E_{ph}$ is the total energy of
gamma photons and $E_{laser}$ is the input laser energy.
(f) Time evolutions of the total number of gamma photons ($N_{ph}$) with energy above $1$ ${\rm MeV}$ and $10$ ${\rm MeV}$ (the inset figure).
The black curves in (e) and (f) show the single target case with the plasma lens only. The blue regions in (d-f) show the period
when the laser reflection occurs.
}\label{fig5}
\end{figure}

The properties of the gamma photons emitted in the nonlinear Compton scattering process are shown in Fig.\ \ref{fig5}. It is shown from Figs.\ \ref{fig5}(a) and (b)
that the gamma photons are bunched with the laser wavelength, which is in accordance with the electron beam shown in Fig.\ \ref{fig3}(b). The duration (FWHM) of the
whole gamma photon beam is about $10$ ${\rm fs}$ (corresponding to $3$ ${\rm \mu m}$), as shown in Fig.\ \ref{fig5}(b). Figure \ref{fig5}(c) shows that the gamma
photons are collimated within a solid angle of $0.18$ steradian (${\rm sr}$), and the maximum number of high-energy gamma photons ($\geq1$ ${\rm MeV}$) emitted per
square degree (or $3\times 10^{-4}$ ${\rm sr}$) reaches about $5.0\times 10^{8}$. Figure \ref{fig5}(d) shows that both the gamma photon number and energy are greatly
enhanced after the laser reflection. Accordingly, the average power of gamma photon emission is also greatly enhanced and the maximum power is
as high as $80$ ${\rm TW}$, as shown in the inset of Fig.\ \ref{fig5}(d). The gamma photon beam has a broad energy spectrum in this strongly nonlinear regime,
as shown in Fig.\ \ref{fig5}(d). The gamma photon energy reaches up to $100$ ${\rm MeV}$, which agrees well with the above theoretical estimation. The spectral
intensity of the gamma photons within $0.1\%$ energy bandwidth (BW) at $1$ ${\rm MeV}$ reaches about $2\times10^{9}$ ${\rm photons/0.1\%BW}$, the
reconstructed duration is about $20$ ${\rm fs}$, the divergence angle is about $400$ ${\rm mrad}$, and the source size is about $2.0$ ${\rm \mu m}$ $\times$ $2.0$ ${\rm \mu m}$,
which gives a peak brilliance of $1.6\times 10^{23}$ ${\rm photons/s/mm^2/mrad^2/0.1\%BW}$ at the source point. In addition, about $2.6\%$ of the
laser energy (or $10\%$ of the electron energy) is converted into the forward-going gamma photon beams, whereby about $2.1\%$ is converted into photons
with energy above $1$ ${\rm MeV}$ and about $1\%$ is converted into photons with energy above $10$ ${\rm MeV}$, as shown in Fig.\ \ref{fig5}(e). As a comparison,
the single target case without the plasma mirror is also considered, as shown in Figs.\ \ref{fig5}(e) and (f). In this case, the nonlinear Compton scattering
cannot occur and the betatron radiation (or synchrotron radiation) by electrons oscillating in the plasma channel becomes dominant \cite{Huang}. It is shown that
in the present scheme using a bilayer plasma device, both the conversion efficiency and high-energy gamma photon number (with energy above $1$ ${\rm MeV}$) are about
one order of magnitude larger than those in the single target case. In particular, approximately two orders of magnitude enhancement can be achieved for the high-energy
photons above $10$ ${\rm MeV}$, as indicated from the inset figures of Figs.\ \ref{fig5}(e) and (f).

\section{Parametric dependence of efficient gamma photon emission}
\label{sec6}
\begin{figure}
\centering
\noindent\makebox[\textwidth][c]{\includegraphics[width=0.8\paperwidth]{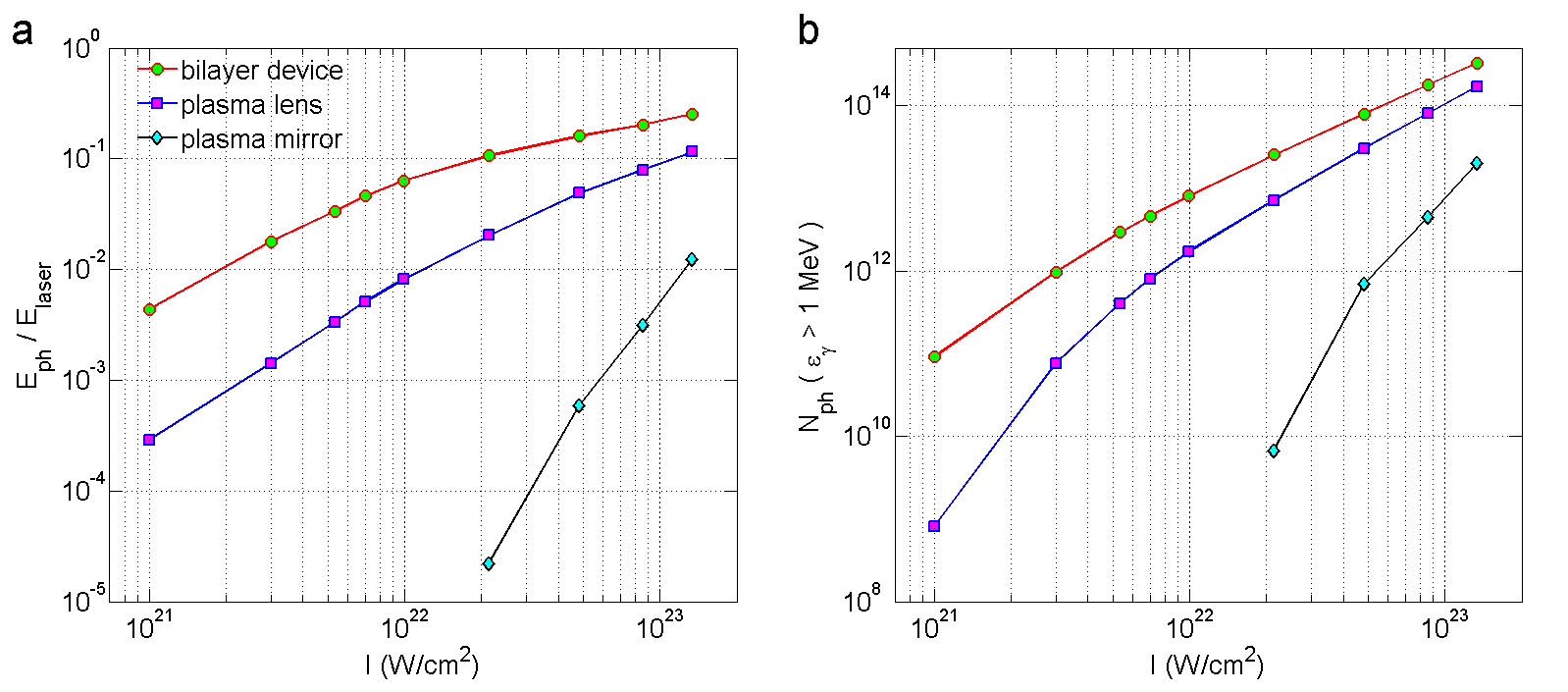}}
\caption{{\bf Intensity scan for different target cases.}
The conversion efficiency for gamma photon emissions (a) and the number of high-energy gamma photons with energy above $1$ ${\rm MeV}$ generated per laser shot (b) at different laser intensities for a fixed parameter of $n_e/a_0n_c=1/50$, where $n_e$ refers to the electron density of the plasma lens. The red, blue, and black curves correspond to the bilayer plasma target case, the plasma lens target case, and the plasma mirror target case, respectively. In the simulations, only the forward-directed photons are calculated. In these cases, the thickness of the plasma lens target are fixed and the parameters of the plasma mirror target are also fixed.
}\label{fig6}
\end{figure}

\begin{figure}
\centering
\noindent\makebox[\textwidth][c]{\includegraphics[width=0.8\paperwidth]{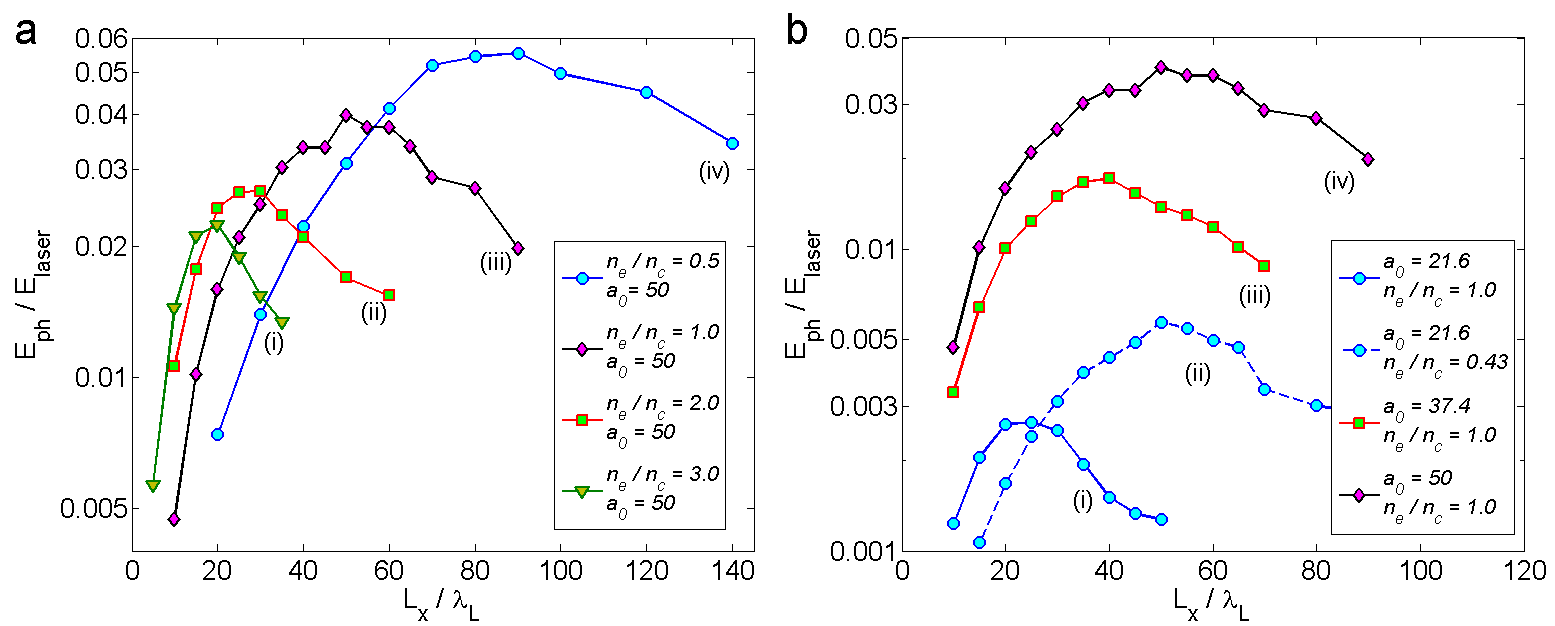}}
\caption{{\bf Dependence of the conversion efficiency on the length of the plasma lens.}
The dependence of the conversion rate of laser energy into gamma photons on the length of the plasma lens ($L_x$) for different plasma densities (a)
and laser intensities (b). The curve (ii) and the curve (iv) in (b) correspond to the same parameter of $n_e/a_0n_c=1/50$, where $n_e$ refers to the electron
density of the plasma lens.
}\label{fig7}
\end{figure}

To demonstrate the robustness and effectiveness of the present scheme for different laser and plasma parameters, a series of two-dimensional
particle-in-cell simulations were conducted. In these simulations, the laser intensity, the electron density, and the thickness of the first
layer target are adjusted but the other laser and target parameters are fixed, which are the same as the three-dimensional simulations. In the
simulations, the simulation box is resolved with the cell size of $8$ $\rm{nm}$ along the laser propagation axis and $20$ $\rm{nm}$ along the transverse axis.
As a comparison, the single target cases either with the plasma lens alone or with the plasma mirror alone are also considered. The corresponding results are
shown in Fig.\ \ref{fig6} and Fig.\ \ref{fig7}. Figure \ref{fig6} shows that the present scheme works effectively for a wide range of laser intensities
and plasma densities. In particular, for the lasers delivering intensities of $I\sim 10^{21}$ ${\rm W/cm^2}$, exploiting the present scheme can achieve
more than one order of magnitude enhancement on the conversion efficiency and gamma photon flux when compared with the plasma lens target case (or the betatron radiation scheme).
The conversion efficiency roughly scales as $I^{3/2}$ for $I\leq10^{22}$ ${\rm W/cm^2}$ in the present scheme. For the lasers at intensities above $10^{22}$ ${\rm W/cm^2}$,
the conversion efficiency can exceed $10\%$. In this case, the radiation reaction effects significantly impact the radiation process and the scaling
of the conversion efficiency is reduced as $I^{1/2}$ \cite{Chang}. In the present scheme, the required laser intensity for efficient gamma photon
emissions can be much lowered, for example, the required laser intensity for $1\%$ conversion efficiency is about $2\times 10^{21}$ ${\rm W/cm^2}$,
which can be easily achieved with the existing petawatt-class lasers \cite{Danson}. In contrast, to achieve $1\%$ conversion efficiency in the single
target cases, the required laser intensity is above $10^{22}$ ${\rm W/cm^2}$ for the NCD plasma lens and is above $10^{23}$ ${\rm W/cm^2}$ for the solid-density
plasma mirror.

In experiments, the length of the plasma lens is also an important parameter to optimize the gamma photon emission. Figure \ref{fig7} shows the dependence
of the conversion efficiency on the length of the plasma lens for different laser intensities and plasma densities. It is shown that there exists an optimal
length of the plasma lens for efficient gamma photon emissions. In the optimal case, the electrons are sufficiently accelerated and the self-focused
(or reflected) laser pulse is sufficiently strong. The optimal length is increased as the plasma density is decreased or the laser intensity is increased,
as shown in Fig.\ \ref{fig7}(a) (see the curves (i), (ii), (iii), and (iv)) and Fig.\ \ref{fig7}(b)(see the curves (i), (iii), and (iv)). In particular,
the optimal length is invariant when the parameter $n_e/a_0n_c$ is fixed, as indicated in Fig.\ \ref{fig7}(b) (see the curve (ii) and the curve (iv)).
It suggests that the optimal length depends on the single parameter $n_e/a_0n_c$, and it can be estimated by the self-focusing length of laser pulse in
plasma lens, i.e., $\sqrt{\ln{2}}\sigma(a_0n_c/n_e)^{1/2}$, where $\sigma$ is the beam radius. For example, for $n_e/a_0n_c=1/50$ the optimal length
obtained from simulations is about $50\lambda_L$, as shown in Fig.\ \ref{fig7}(b), which is in good agreement with the theoretical self-focusing length ($51\lambda_L$).
When the plasma length is less than this optimal length, the laser-shaping process is less efficient and the energy converted into the electrons is also low,
thus the Compton scattering process is less efficient. When the plasma length is larger than the optimal length, the laser pulse is seriously depleted and then
the reflected laser energy is limited. Furthermore, in order to enable the laser reflection process and thus the Compton scattering process, the length of the
plasma lens should be less than the laser depletion length in NCD plasmas, which can be approximated as \cite{Huang} $c\tau n_ca_0/4n_e$, with $\tau$ being laser
pulse duration.

\section{Discussions}
\label{sec7}
The present scheme using a microsized bilayer plasma device provides a compact and efficient way for producing high quality $\gamma$-rays.
For the available lasers at intensities $\sim 10^{21}$ ${\rm W/cm^2}$, brilliant ($\sim 10^{23}$ ${\rm photons/s/mm^2/mrad^2/0.1\% BW}$)
$\gamma$-rays with very high conversion efficiency ($10^{-2}$-$10^{-1}$) and spectral intensity ($10^{9}$-$10^{10}$ ${\rm photons/0.1\%BW}$)
can be realized in the present scheme. Such $\gamma$-ray source is about $10^5$ times more brilliant than a laser-driven bremsstrahlung
source \cite{Kmetec,Gahn,Edwards,Glinec,Giulietti,Cipiccia} and a conventional Compton $\gamma-$ray source based on large-scale
accelerators \cite{Albert2010}. When compared with the recently reported Compton $\gamma-$ray sources based on laser-wakefield
electron accelerators \cite{Schwoerer,Powers,Chen,Sarri,Khrennikov,Yan,Phuoc,Tsai,Dopp,Yu}, the present $\gamma$-ray source is about ten times more brilliant,
and especially its spectral intensity is about $10^4$ times higher. In particular, in contrast to the previous Compton $\gamma$-ray sources
with a conversion efficiency less than $10^{-5}$, the present scheme can deliver an unprecedentedly high conversion efficiency. In addition,
in comparison with other existing schemes (e.g., the betatron radiation or synchrotron radiation) using a single plasma target, exploiting the current
scheme also makes it feasible to produce $\gamma$-rays with much superior properties in terms of conversion efficiency, spectral intensity,
peak brilliance, and photon flux.

The efficient and intense gamma-ray source generated in our scheme could be useful for a wide range of applications. In particular, many
of applications for photonuclear spectroscopy, inspection of nuclear waste, material synthesis, cancer therapy, etc., require sufficient
integrated dose in as a short time or a small number of shots as possible. The gamma-ray emitter proposed here promises a superiority in
spectral intensity and a micrometer-sized monolithic structure to the existing high-energy accelerator-based laser Compton sources,
which have four to five orders magnitude lower spectral intensity and a $100$-${\rm m}$ size footprint. Such high-flux gamma-ray source
here operated at the repetition rate of $1$ ${\rm Hz}$ is comparable with a conventional low-flux source operated at the repetition rate
of $10-100$ ${\rm kHz}$. The recent advance of petawatt laser technology already makes it possible to be operated at $1$ ${\rm Hz}$ \cite{Danson}.
As an example, the high-flux gamma-ray source presented here could be useful for nuclear spectroscopy such as nuclear resonance fluorescence,
by observing the fluorescence at a large angle from the initial photon beam direction \cite{Bertozzi} or by narrowing the bandwidth with a
spatial filter \cite{Albert2010b}.

The present scheme can operate for a wide range of laser and plasma parameters. To realize this scheme in experiments, the
key is to employ the NCD plasma lens that is relativistically transparent for a high-intensity laser pulse. Recently several techniques
such as highly compressed gas jet \cite{Helle}, ultra-low density plastic foams \cite{Chen2016}, and carbon nanotube foams \cite{Bin}, have
been proposed to produce NCD plasmas, which makes it feasible to achieve the present scheme within current technical capabilities.
In addition, in order to avoid destroying the plasma lens by the pre-pulse that deteriorates the temporal quality of the laser pulse,
a high contrast $10^{-10}$ for the high power laser of intensity $10^{21}$ ${\rm W/cm^2}$ is necessary in experiments because in such
case the pre-pulse cannot create pre-plasmas, which was demonstrated in recent pre-plasma experiment using the ultrafast optical parametric
amplifier petawatt laser \cite{Wagner}. It is noted that in previous Compton scattering experiments using a plasma mirror, the pre-pluses lead
to significant reflectivity reduction of the plasma mirror \cite{Tsai2}, which will reduce the efficiency on the Compton scattering process and
thus the gamma-ray production. In our proposed scheme, such an issue can be also suppressed due to the fact that during the laser-shaping process
in plasma lens, the pre-pulses can be effectively absorbed and the pulse contrast can be significantly improved prior to the plasma mirror \cite{Wang,Bin},
which makes the effects of pre-pulse become negligible in our scheme. Experimentally, an ultra-intense laser pulse of the order
of $10^{21}$ ${\rm W/cm^2}$ with the contrast level better than $10^{-10}$ and even $10^{-11}$ at $6$ ${\rm ps}$ prior to the peak has been
achieved by using a double plasma mirror system \cite{Kim}, which can greatly reduce the level of pre-plasma creation. Accordingly, the experimental
feasibility of our proposed scheme is quite convincing for the state-of-art petawatt laser system.

Bremsstrahlung radiation can be produced when the electron beam accelerated in the plasma lens interacts with the plasma mirror target. However, the bremsstrahlung
radiation only becomes effective for electrons interacting with a thick target composed of high-Z materials. In our scheme, the number of photons with the energy from
$E_{min}=0.01$ $m_ec^2$ to $E_{max}=100$ $m_ec^2$ via bremsstrahlung radiation generated by electrons with energy $E_e=1000$ $m_ec^2$ and charge $Q_e\sim 10$ ${\rm nC}$
interacting with a 1 ${\rm \mu m}$ thick aluminum plasma mirror target can be estimated as \cite{Koch} $N_{\gamma}=1.97\times 10^{4}\left(\frac{Q_e}{1{\rm nC}}\right)\ln\left(\frac{E_{max}}{E_{min}}\right)\left(\ln(\frac{2E_e^2}{\sqrt{E_{min}E_{max}}})-\frac{1}{2}\right)\approx 2.54\times10^{7}$. To confirm the
above estimation, a simulation of bremsstrahlung radiation using the GEANT4 code \cite{Jeon} is also conducted, which gives a total number of gamma photons
about $1.75\times 10^7$, on the same order of our theoretical estimation. Hence, the number of bremsstrahlung photons generated in aluminum plasma mirror is
negligibly small, compared to the number of photons of the order of $10^{12}$ (see Fig.\ \ref{fig5}(f)) generated via nonlinear Compton scattering. Thus the
contribution of bremsstrahlung radiation can be neglected in our scheme.

\begin{figure}
\centering
\noindent\makebox[\textwidth][c]{\includegraphics[width=0.8\paperwidth]{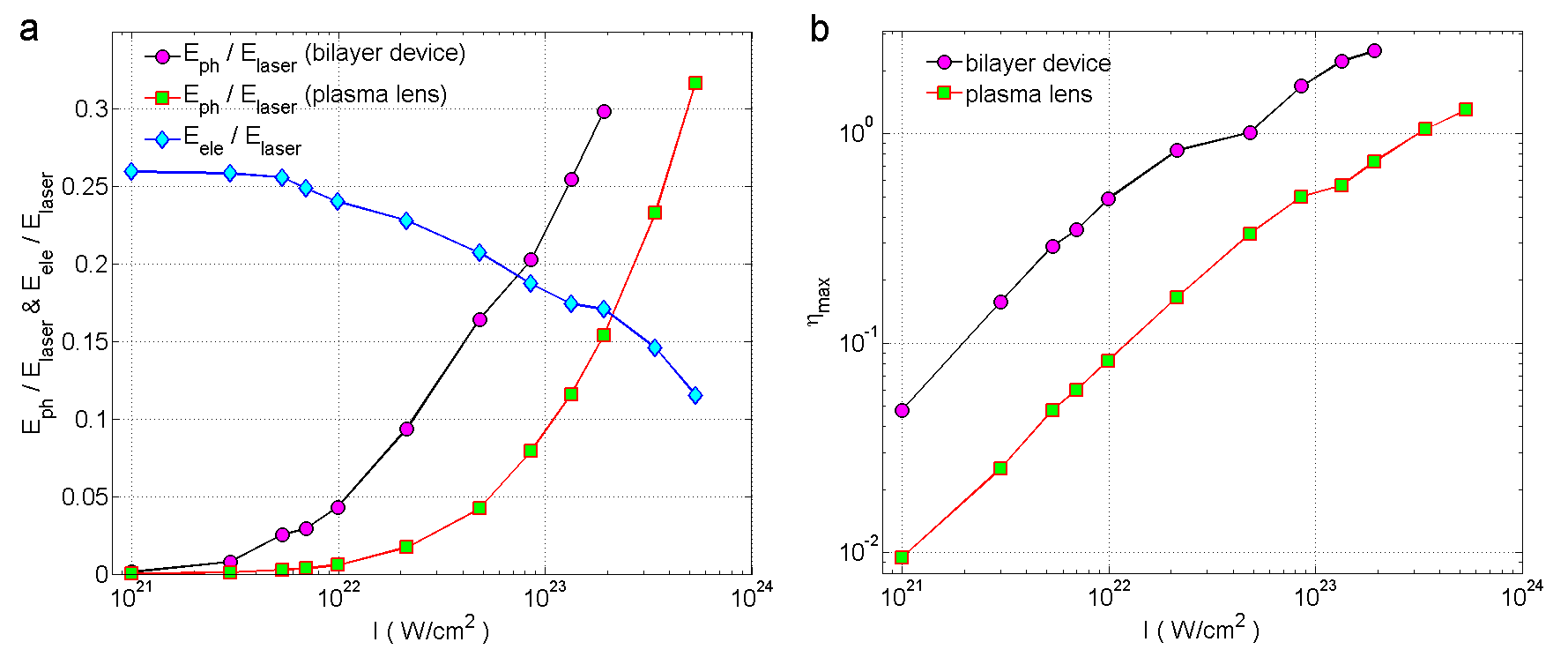}}
\caption{{\bf Intensity scan on the conversion efficiency and the quantum emission parameter $\eta$.}
(a) The conversion rate of laser energy into the gamma photon beam and the electron beam at different laser intensities.
(b) The corresponding maximum value of $\eta$ for electrons at different laser intensities. The black curve and red curve in (a) and (b) show the
case with bilayer plasma target and the case with only the plasma lens, respectively.
}\label{fig8}
\end{figure}

The robustness of the present scheme makes excellent prospects for next-generation lasers. For example, the required laser intensity for reaching
the quantum-radiation-reaction dominated interaction process (corresponding to $\eta \geq 1$) can be much lowered in the present scheme, as shown in
Figs.\ \ref{fig8}(a) and (b). This makes it possible to explore the quantum-radiation-reaction effect, which is an intriguing unsolved problem of quantum
electrodynamics, in laboratories by employing a laser with intensity of $\sim 10^{22}$ ${\rm W/cm^2}$. In this case, gamma photon emissions become the
dominant energy absorption process and their energy even exceeds the accelerated electrons, as indicated in Fig.\ \ref{fig8}(a). However, the required
laser intensity for reaching this regime is above $10^{23}$ ${\rm W/cm^2}$ in the plasma lens case (or the betatron radiation scheme), as shown in Fig.\ \ref{fig8}(b).
In addition, in strong laser fields, electron-positron pairs can be produced when gamma photons interact with the laser photons via the multi-photon
Breit-Wheeler process. In the head-on colliding configuration presented here, the quantum parameter that determines the probability of pair production
can be expressed as $\chi\approx (\varepsilon_{\gamma}/m_ec^2)E_{\perp}/E_s$ \cite{Ridgers,Duclous}, where $\varepsilon_{\gamma}\approx \eta\gamma_em_ec^2/2$ is the gamma photon
energy, $E_{\perp}$ is the electric field perpendicular to the photon momentum, and $E_s$ is the Schwinger field. In this case, one has $\chi\approx\eta^2/4$.
When $\chi\geq1$, the Breit-Wheeler process becomes dominant and pair production plays a significant role. It is shown from Fig.\ \ref{fig8}(b) that pair production
would become dominant ($\chi\geq1$ or $\eta\geq2$) when the input laser intensity exceeds $10^{23}$ ${\rm W/cm^2}$. For future lasers at intensities
above $10^{23}$ ${\rm W/cm^2}$, exploiting this scheme could open up the possibility of creating high-flux ${\rm GeV}$ photons \cite{Gong} and dense electron-positron
pairs \cite{Zhu}, which are especially attractive for laboratory astrophysics \cite{Bulanov}.

\section{Summary}
\label{sec8}
In summary, we have proposed a highly efficient and compact laser-driven gamma photon emitter, which can convert an appreciable fraction
of laser energy into high-energy gamma photons for a laser with intensity of $\sim 10^{21}$ ${\rm W/cm^2}$.
This novel gamma photon emitter consists of a plasma lens and a plasma mirror irradiated by a single laser pulse.
The present scheme harnesses the nonlinear Compton scattering process through the combination of relativistic plasma optics
and the DLA electron accelerator that produces extremely high-charge, high-energy electron beam. This scheme opens up the possibility
for developing ultra-brilliant $\gamma$-ray sources with unprecedentedly high conversion efficiency and spectral intensity, which holds great
promise for many applications in a broad range of fields, including medicine, industry, military science, material science, and high energy density science.

\section*{Acknowledgements}
We acknowledge the fruitful discussions with Prashant Kumar Singh, Vishwa Bandhu Pathak, Naser Ahmadiniaz, and Ki Hong Pae and we also acknowledge
Jong Ho Jeon for the help provided in GEANT4 simulations. This work was supported by IBS(Institute for Basic Science) under IBS-R012-D1, and also by GIST
through the grant ``Research on Advanced Optical Science and Technology". The EPOCH code was developed as part of the UK EPSRC grants EP/G056803/1
and EP/G055165/1.

\end{document}